# Guiding Development Work Across a Software Ecosystem by Visualizing Usage Data

**Christopher Bogart**
Carnegie Mellon University
Pittsburgh, PA USA
cbogart@cs.cmu.edu

**James Howison**
University of Texas at Austin
Austin, TX USA
jhowison@ischool.utexas.edu

**James Herbsleb**
Carnegie Mellon University
Pittsburgh, PA USA
jdh@cs.cmu.edu

**ABSTRACT**
Software is increasingly produced in the form of *ecosystems*, collections of interdependent components maintained by a distributed community. These ecosystems act as network organizations, not markets, and thus often lack actionable price-like signals about how the software is used and what impact it has. We introduce a tool, the Scientific Software Network Map, that collects and displays summarized usage data tailored to the needs of actors in software ecosystems. We performed a contextualized walkthrough of the Map with producers and stewards in six scientific software ecosystems that use the R language. We found that they work to maximize diversity rather than quantity of uses, and to minimize coordination costs. We also found that summarized usage data would be useful for justifying ecosystem work to funding agencies; and we discovered a variety of more granular usage needs that would help in adding or maintaining features.

**Author Keywords**
Software ecosystems; Scientific software

**INTRODUCTION**
Software ecosystems are collections of interdependent components maintained by a distributed community. Ecosystems are an increasingly important way of producing software, but they inherently fail to provide participants with key information they need in order to decide how to allocate their effort. The value of maintaining these ecosystems is clear: ecosystems such as CRAN (the Comprehensive R Archive Network), Eclipse, Android, and Node.js provide resources that facilitate software development work, allowing developers to use existing software components, libraries, and frameworks developed and maintained by others. Software components are combined and extended to produce innovative functionality, yet the components are built and maintained by a large and diverse population of individuals, organizations, and communities [10].

Ecosystems are "network" organizations [29], lacking the hierarchy of a firm, and eschewing explicit price-based transactions. There is typically no centralized authoritative decision-making about where ecosystem members should spend their effort, and neither is there the classic market signal of price to let producers know where the greatest value can be created. Although network organizations transmit enormously rich information between neighbors in the network, this information is not summarized in an actionable way for questions of global scope for the network. Thus aside from those few products that become widely known and deployed, developers have very little information about if and how their code is used, and are often surprised to find it is used more widely, by more people, and in different ways, than they realized [7].

In software ecosystems, developers typically write software that they themselves need, or that the companies that employ them need [21]. Participants are often willing to do extra work to turn the software they wrote for themselves into a resource the community can use [33], but are reluctant to do so unless the community needs are clear and demonstrable. They have rather limited information about the requirements of the larger community, however, typically in the form of bug reports, feature requests, comments on mailing lists or social media [12], and perhaps work others are doing to modify forked copies of their code [21]. Generally a very small proportion of potential users contribute information in these channels, and surges of attention often represent an insider controversy rather than a reflection of widespread need [34].

Communities of scientists who share software provide particularly compelling examples of this information gap (e.g. [24,33]). Understanding and assuring compatibility and interoperability across these ecosystems presents a substantial information and coordination challenge (e.g., [7,9,38]). Science increasingly depends on software for analysis, modeling, visualization, and storing and manipulating data. Yet resources for developing software are generally very scarce, so making good decisions about effort allocation is particularly critical. Therefore, we selected a set of related scientific software communities in order to address the question: what can computer support systems offer to play the informational role that prices play in markets, to help align effort allocation with needs in software ecosystems?

We developed visualization tools – the Scientific Software Network Map ("the Map") to allow scientists developing software to answer key questions about how much their software was used, and what other software it was used with. The Map is designed to provide meaningful signals about whether maintaining and enhancing specific packages is worth the effort, and about potential interoperability issues with other packages it is commonly used with.

We evaluated the tool's potential usefulness by first asking scientist-developers who write programs in the R statistical language to reflect on their current information-seeking practices. We then provided an instance of the tool using data from the R ecosystem, and allowed them to interact with it to answer any questions they might have. We asked them to evaluate the tool with respect to their own needs. Their responses allowed us to evaluate the capabilities of the tool, as well as our underlying assumption that the scientists are trying to maximize use of their software and minimize their development effort.

Our interviews showed that scientists' behaviors differed in some ways from what our simple market and price metaphor would suggest. The preference revealed by their behaviors was not for putting a greater quantity of better-integrated packages in more hands for the least development effort, but instead for providing for a greater quantity of distinct *use cases*, with the least *coordination effort*. Consistent with the preference for more use cases, they are primarily motivated to program for their own needs, or those of other researchers that are distinctive enough to possibly yield new collaborations or citations. Their approach is mostly reactive: scientists respond to their own needs or the needs of colleagues that draw their attention, but they are mostly not motivated to proactively research the "market" of potential users of scientific software to provide the greatest good for the greatest number. Our evaluation suggested tool modifications to fit the needs of this ecosystem, i.e., a data collection and visualization tool should highlight variety of uses to incentivize development, and give more focused help with inter-project dependencies to lower the cost of coordination.

## RELATED WORK

### Software Ecosystems

Lungu et al. [25] define a *software ecosystem* as: "a collection of software projects which are developed and evolve together in the same environment. The environment is usually a large company, an open-source community, or a research group". Other definitions (e.g. [27]) add the relationships among the developers of those projects as part of the core definition, but both perspectives convey the idea of distributed actors collaborating with each other to build and maintain software projects that rely and depend on each other.

A few studies exist that examine software assemblages and the human infrastructures supporting them as ecosystems [15,16,25,28,35]. These studies stress that software development is making an important shift from standalone applications to ecosystems, where components within an ecosystem work together as a platform for further construction.

In a *scientific* software ecosystem, many scientists, who are primarily engaged in their scientific work, are also creating and maintaining software. Communities of scientists are migrating to such ecosystems, adopting a variety of names, including cyberinfrastructure, grid computing, collaboratories, and eScience [2,23].

Lee, Bietz, and Ribes [23] describe how, in cyberinfrastructure, requirements tend to evolve rapidly in response to new technologies and scientists' diverse needs [23]. Bietz and Lee [3] explored the tradeoffs in the way these systems are adapted with work-arounds, from-scratch development, and extending existing cyberinfrastructures.

Because the incentives for software sustainability in science can be missing or indirect [15,16], scientific software in some fields is characterized by redundant implementations of large monolithic codebases [5, 16], poor support, and infrequent maintenance. These effects come about for several reasons, including heterogeneous needs and timing of different research projects [17], the tension between long and short term needs [30], and reluctance to be dependent on outside parties [6].

### Incentives, Visualization, and Impact

Scientists are rewarded for scientific impact, measured (imperfectly) by things like publication counts and citation of papers. Software plays a large and expanding role in enabling science, but it is cited haphazardly in scientific writing [13,19]. For this reason, the scientific impact of the work of developing and maintaining a piece of software is often invisible to the community [16]. If the use of a software package were made visible along with its impact in the form of publications it enabled, scientists could hope to receive credit for the scientific impact of their software work. Recognition could provide a powerful incentive to do the extra work to make software useful to a broader community, just as recognition has done in online communities [20].

Besides this extrinsic motivation, it has also been observed [33] that knowing what other scientists need is intrinsically motivating to them: they share their software precisely because they perceive that others need it. Usage data about scientific software has the potential to demonstrate to an author that they do have users who need their software, and this should motivate them to continue supporting and enhancing the functionalities that are most used. Knowing how and how much a package is used is very helpful in deciding what work is most worthwhile.

In this paper we evaluate a tool aimed at two primary types of users [14] within scientific software ecosystems: the software *producer* who writes software that could potentially be used by others, and an *ecosystem steward*, who is anyone concerned with the health and unity of the set of packages as a whole, and how well they meet the needs of a scientific community.

**SCISOFT NETWORK MAP TOOL DESIGN**

The Map is designed to be populated from different ecosystems' software repositories. The interface uses d3 for the visualizations, and pyramid, mongo and jinja for the web and database framework.  Maps are designed to directly address the needs of scientific software *producers* and *stewards* for usage-related information about packages. The tool's features include a usage graph over time, a filterable/sortable list of packages, a "co-usage" graph showing what packages were used together, and a listing of external software (e.g. end-user scripts and packages under development) that depend on each package.

The Map frames software contribution in terms of the positive impact it was having on others; we intended it to *motivate* scientists to further this end; and in practice provide numbers and graphs that scientists could show to tenure committees or granting agencies in order to *justify* their work; it can also *inform* them of usage patterns to help direct and prioritize development.  Visualizations in the tool include:

PACKAGELIST: The main directory of packages emphasizes packages' importance by usage or impact, by ordering packages by user-selectable measures of impact, and filtering them by ecosystem. The measures include counts of uses, "recent" uses, distinct users, publication counts, and usage counts among users' publically shared projects. Scientific publication counts come from Scopus [31]. Besides simply serving as a directory and entry point to the tool, this listing is designed to draw scientists' attention to the most-used options in a fragmented field, centralizing attention and resources to provide better economies of scale and promote standardization. (via, e.g. rational herding [8]). The idea is to help align the incentives in the ecosystem, encouraging technical work on software that is actually used.

CITELINKS: Links to the actual papers are available via Scopus. The intention is that producers could refer to these in a CV to show the impact of their work, or they could use it to channel their efforts into their most-cited packages. Stewards could use it to show the overall benefit of the ecosystem to grantors. It could also help producers see which of their colleagues are citing different packages, and read the papers to see how they are being used.

USEHISTORY (Figure 1(b)): The Map can also depict usage history over time for a user-selected set of packages, showing how many users were using different versions of the software at each point in time. This visualization is designed to help *producers* track trends in their package's use, and thus demonstrate to others that their ecosystem contributions are being used; it can help *stewards* maintaining sets of shared packages spot trends such as emerging use of new packages or shifting adoption of new versions. It could also help them make decisions about allocating resources and making packages available.

**(a) CoUsage**

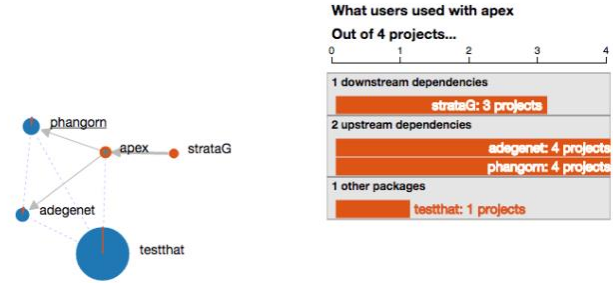

**(b) UseHistory**

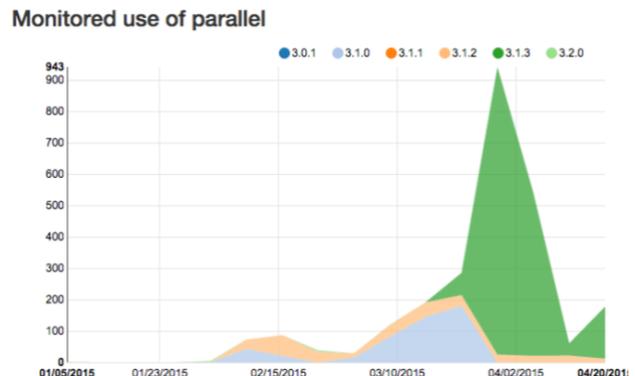

Figure 1. Two features of the Map: (a) *co-usage* visualization of software artifacts (nodes), frequency of use (node size), frequency of use together (edge width), type of dependency (dotted = logical; solid = formal), and relevance to the focal package of the diagram. (orange = only used with apex; blue = also used when apex not used).  The bar graph at right shows the data in another form: bars show how many projects using the focal package also used some other package. (b) Number of runs of each version of a package over time.

USERPROJECTS: The tool lists projects that rely on each package.  We designed this facility to allow *producers* to see how their package is being used by end users, not just by the other packages in the ecosystem, and drill down to the level of specific lines of source code that call their package's API, on the theory that being aware of specific end-user usage patterns might prompt producers to put more time into developing the more popular packages or parts of their packages.  It could also allow end users to find concrete examples of how other users may have used these packages.

COUSAGE: This feature (Figure 1(a)) depicts which packages were most commonly used *together*, on the theory that software integration/compatibility work should be considered as a valuable scientific contribution: if scientific end users often find themselves trying to integrate the functions of two packages, then producers of those packages could potentially assist in the work of those end user scientists by making their packages work together more easily, adding things like data structure conversions, documentation relevant to those circumstances, or code that adheres to standards typical of that environment. *Stewards* can also see which of their packages merit better interoperability efforts, and end users can get ideas for useful software combinations from their colleagues.

This view depicts the neighborhood of a single package in the graph, and allows navigation to neighboring packages in the graph. It shows *static* dependencies (packages that require others as prerequisites), and *logical* dependencies (packages that users chose to use together to solve some problem). To keep the graph readable, links with low pointwise mutual information are pruned: so that, for example, a package G that is very popular, and whose use is uncorrelated with the use of package P, is unlikely to be shown in the CoUsage graph's for package P.

*Data presented by the Map*
R is a computer language specialized for statistical computing. A user starts it up from the command line or by running a graphical environment like RStudio [1]. From there, they can run scripts that they or others have written, or type commands directly at a prompt. The scripts or commands may in part rely on functionality provided by *packages*. R has a rich ecosystem of many thousands of specialist plugin packages for advanced statistical techniques, machine learning, and domain-specific areas from bioinformatics to fishery science. Packages are units of software functionality that are typically written by scientists who need them for their own work, freely shared through websites like cran.org and bioconductor.org, and are easy for users to install. There are sub-ecosystems of packages, nested within the larger R ecosystem, that manage their own interdependent collections of related packages.

We created an instance of the Map with data from three open software repositories containing R software (CRAN, Bioconductor, and Github, using the mirror GHTorrent [11]). We are collecting data on an ongoing basis of the package usage of over 100,000 R-language end-user scripts and other projects shared on Github, a platform for sharing source code of open-source software. Map downloads the code for new or changed projects once a day, and examines which packages they were using.

Some programmers use Github to store their source code even for small projects whose adoption they may not be actively trying to promote, because of its convenience and a principled desire to work openly. It thus captures a range of uses that overlap with scientific usage. As we shall see below, stewards had different takes on the appropriateness of Github data depending on their ecosystem.

To link academic citations with packages, we read a metadata field that many R packages provide that suggests a canonical citation for users to refer to. We look up citation counts to these papers in Scopus [31] on a rotating basis, approximately every two weeks.

## STUDY
The Map is available as an online web application (http://omitted.for.anonymity), with more than 2000 users

**Impact**
What do you do to estimate usage and impact?
What do you do with that information?

**Users and their needs**
How do you find out and prioritize what needs work?

**Coherence/Co-usage**
What work do you do towards interoperability of packages?
Do you reject or prune packages based on duplicate functionality? (Stewards only)
How do you find out what interoperability issues need work?

**Evaluation**
How would you use the information presented in the tool?
Does it seem correct?
What is missing that would make it more useful?

**Figure 2. Semi-structured interview topics for package producers and ecosystem stewards**

(as measured by Google Analytics) since its introduction earlier in 2015. It has been instantiated for two ecosystems: the R statistical language, and the ecosystem of supercomputer applications available at the Texas Advanced Computing Center (TACC).

In this study we evaluate the Map in the context of the R ecosystem, asking producers and stewards in sub-ecosystems of R about their current practices, and evaluating in a walkthrough how the tool could help them. We address the following research questions:

RQ1 Did usage and impact information help to *motivate* scientists to do the work of scientific software construction and integration, and help them *justify* this work to the decision-makers they answer to?

RQ2 Did software producers and stewards consult usage information to weigh cost/benefit decisions about what software work to do and how to prioritize it?

RQ3 Did information presented in the tool fit interviewees' mental models of their ecosystems? What do any differences imply about how usage data should be collected and presented?

## METHOD
We sought out interviews with people involved in R ecosystems who were directly responsible for adding, removing, and maintaining software. We operationally defined "ecosystem" as a set of packages for which there was evidence of a common purpose, and of an organization or community, with a website, advocating for adoption or interoperation of the packages.

In these ecosystems, we interviewed package *producers* (identified as people listed as the maintenance contact for a published R package) and *ecosystem stewards* (identified as individuals listed as contacts on the ecosystem's web page). In the organizations we considered, *ecosystem stewards* were also producers since they were writing code, both for their own scientific reasons, but also in promotion of standard data structures or interfaces within the ecosystem.

| Ecosystem | Purpose | Interviewees |
|---|---|---|
| Bioconductor www.bioconductor.org | Bioinformatics | 2 Stewards (S-Bio-1,2) |
| FLR www.flr-project.org | Fishery management | 1 Steward (S-FLR-1) |
| rOpenSci ropensci.org | Open data, reproducible science | 1 Steward (S-Sci-1) |
| rOpenGov ropengov.github.io | Government data | 1 Steward (S-Gov-1) |
| (no name) github.com/NESCent/r-popgen-hackathon | Population Genetics | 5 Producers (P-Gen-1,2,3,4,5) |
| CRAN cran.r-project.org | General purpose | 2 Producers (P-Cran-1,2) |

**Table 1. The twelve interviewees. Participants' codes represent their role and ecosystem affiliation**

We did not attempt to interview users or people with other roles in the ecosystems, since they contribute to the software only indirectly.

We interviewed 12 people, listed in Table 1. The interview was designed to evaluate the software by a *contextualized walkthrough*: we first asked them about their current practices, then walked them through the screens of the tool, showing them displays of their own packages' usage, impact, and relationship with the ecosystem, and asked them if it could replace or augment their practices. We followed a semi-structured interview format, with questions (Figure 2) derived from the list of information needs for collaborators in software ecosystems as described in the related work section and in Howison's enumeration of information needs [14]. Questions about the tool focused on the intersection of issues that came up in the current practice part of the interview, and ecosystem data viewed via the Map.

We wanted to interview producers whom we thought were likely to have encountered different kinds of issues where information about use is important. We selected five interviewees from attendees of the NEScent Population Genetics hackathon [22], which was in part aimed at improving interoperability among R packages for phylogenetics and related fields, since understanding use is key to interoperability. We also interviewed two producers of different general-purpose packages, with large and varied user communities who could be using the packages for highly varied purposes.

We also interviewed five ecosystem stewards, individuals listed as having central roles on the websites of four other R ecosystems of varied sizes and domains: Bioconductor (a very large set of biology-related packages, with 1023 software packages, plus 1122 data packages); FLR (for evaluation of fishery modeling and management strategies with 9 packages), rOpenSci (which facilitates open data and reproducible research with 87 packages), and rOpenGov (which facilitates access to government data sources with 32 packages).

**RESULTS**

This section is organized around the three research questions: (RQ1) motivating and justifying ecosystem development work in general, (RQ2) determining and prioritizing what particular work to do, and (RQ3) tool differences from users' mental models.

**Justifying and Motivating Development**

**Producers' practices** Producers needed data to *justify* their work to decision-makers, but the specifics varied. They did not mention sources of data that were *motivating* to them, but instead described abstract motivations such as reciprocity, scientific accuracy, and helping other users.

Producers were varied and unsystematic in what information they attended to and used to justify their work on software. Some of them said they needed to provide evidence that their work had impact, but the evidence they currently rely on varied, and none of them expressed doubt that they would be able to satisfy this need with their current practices.

Some producers mentioned tracking citations to their software or to a related paper they asked users to cite:

*At some point, within this next release, I'm hoping to have some sort of publication announcement, a vignette or something, that I can point to as a citable reference. And that will be a bit of a justification as well. [P-Gen-1]*

*So, this is part of my thesis, and I do have to justify it, but it is published [as a methods paper], and methods papers are always highly cited. So I can justify it by saying, "This will always be my highest cited paper," because it's already gotten 14 citations within the past year, and it's only been out for a year. [P-Gen-4]*

One package author said that other packages came to rely on his package (a relationship which is easy to see in CRAN), and this had helped get him his current job:

*I think the post doc here I got more or less because I have this package out. [It] has ten packages which depend on it. So, that's kind of a sign that it's important. [P-Gen-3]*

Producers, in short, relied on whatever information about impact was available and salient to justify their work but were not particularly eager to find better sources of information.

Producers' impact on others' scientific efforts was also personally *motivating* for them. Their reasons were abstract, focusing on helpfulness, correctness and reciprocity, rather than mentioning particular metrics as they did when discussing justifications.

One described the reciprocity of sharing as motivating regardless of whether it was actually helpful:

> *To be honest, I do not really think about whether it is helpful or not. It is certainly helpful to the work I am doing, and I was happy to share this [...] in the same spirit as others share their work/packages in the R community. [P-Cran-2]*

Another spoke vaguely of what "might be helpful", but did not elaborate on ways of verifying that the code was really helpful:

> *I thought, "Hmm, well, then, perhaps others might find this helpful too, and it's yet another thing which will be out there," and I made it available. [P-Cran-1]*

A third was more specific about what might be helpful: improving others' results:

> *They're using methods that are giving them the wrong answer. We think that our method will give them better answers will give more correct answers. So, we want the answers that people publish to be right. [P-Gen-2]*

**Producers' evaluation** The USEHISTORY, PACKAGELIST, and CITELINKS features of were designed to help demonstrate the scientific impact of software, and producers' reactions to it suggests that they perceived the kinds of data provided as useful, with some caveats about the details of our implementation. In the Map walkthrough, we started off most interviewees by showing them the PACKAGELIST, filtered to their ecosystem and sorted by one proxy for "usage": the number of Github projects which referred to the package.

Interviewees were accustomed to using citation counts to justify their work, and liked the convenience of having them listed for each package. However they were almost universally critical of our choice to use Scopus counts in contrast to Google Scholar. Google scholar seems to be an approximation that interviewees preferred: it casts a wider net and errs on the side of false positives rather than false negatives. We know this because the critiques were quite pointed: when Scopus was unable to recognize a citation in order to count references to it, authors expressed concern about their software getting short shrift:

> *It's a bit strange...I know of many papers which cite my package (the software documentation) so it should really count them (but Scopus seems not to). [P-Cran-2]*

Unfortunately, Google Scholar's terms of service prohibit automated search for citations.

As for *motivation,* one producer stated a direct motivation from evidence of surprisingly broad co-usage to do improvement work on his package:

> *I'm really getting nervous when I see that so many people use that package. [Laughter] That should really motivate me to revisit the code and actually make it numerically more stable [P-Cran-1]*

Producers commented on the convenience of several features that were consistent with their abstract motivations: high usage of a package could suggest to an author that it has been helpful or that the act of reciprocity has been accepted by a community; and the ability to find and examine uses of one's package could help reassure an author that people are improving their work by correctly employing a package's features.

**Stewards' Practices** Ecosystem stewards drew from more varied information sources about impact than producers did, and put more effort into synthesizing the information into coherent reports for granting agencies and other decision-makers.

S-FLR-1, for example, said their organization tries to estimate counts of individual and organizational users. They count organizations that published "grey literature" that mentioned FLR software, they count attendees and their sponsoring organizations at workshops, and tally personal contacts to produce numerical estimates of numbers of attendees.

Other stewards did not try to produce unified estimates of number of users, but instead drew on many sources to produce reports and proposals to granting agencies; making graphs, charts and tables summarizing data that was evidence of widespread usage, interest, and scientific impact of ecosystem packages. Measurements they mentioned included number of attendees at workshops and training classes they had sponsored, numbers of citations for popular packages in Google Scholar; number of results returned from keyword searches in academic and grey literature; volume of activity in discussion boards, mailing lists, and twitter mentions; number of emails and personal contacts at conferences; package download counts, and website visits. In the Discussion section we suggest how the Map might be extended to help with some of this data gathering.

**Stewards Evaluation** Stewards generally liked the idea of having this kind of information collected together in one place:

> *This is the kind of thing we need to think about to be fairly honest. [S-Flr-1]*

> *I think it's important to know how [the packages] are used. Especially when we are developing and maintaining multiple packages, it's really good to have this kind of overview. And if you have 10 or 20 or more packages, what of them are the most dominantly used? And I'm sure it can easily happen when some packages don't really find users and some might become very popular. And I think it's quite useful to be able to distinguish them according to the usage. [S-Gov-1]*

We did not show USEHISTORY to all the users, but one steward showed interest in the data, commenting that they had not pursued similar data that was already available to them, but would have taken some work to extract:

> *We can recover the download stats [but] I don't think we have ever really looked at it. It would be interesting to know for us [S-Bio-1]*

**Choosing development needs and setting standards**

*Community needs*

**Producers' practices** We asked producers how they knew what features to add to their packages, and the practices they described were mostly reactive: that is, they worked on new use cases that came to their attention rather than proactively seeking out user needs for new or improved features. New use cases came primarily from their own research needs, but also from user requests that struck them as easy, interesting, or fruitful, and from incompatibilities that they scrambled to fix when other neighboring packages were updated.

> *Use cases mostly came from our own research requirements. [S-Bio-2]*

> *I developed for myself because it didn't exist, and I figured it was a general enough issue that other people could use it. [P-Gen-1]*

Interviewees agreed that priorities for the ecosystem as a whole was a broad diversity of uses, driven by the individual package authors meeting their own research needs:

> *It's very weakly defined by a central vision, and much more so defined by the agendas of the individual contributors. [S-Bio-2]*

But besides providing for their own research needs, producers did describe ways that the needs of others fit into their practices, mostly in response to direct, usually time-sensitive requests from other ecosystem actors or automated tools.

The CRAN repository has semi-automated mechanisms for alerting package authors when a change in a neighboring package, or R itself, causes an incompatibility. One producer told us this was almost the only time he updated his 20-year old package now:

> *R itself changes so that packages don't work anymore: there are these quality-control tests that the R maintainers have imposed, and then you are forced to do some changes. [P-Cran-1]*

CRAN's policies require some trust between producers. Their policy is to "archive" any package that does not respond quickly enough to an email from the CRAN team saying that it has failed a test [26]; so if a producer chooses to rely on another package, and that other package's maintainer falls behind on maintenance, the producer's own package may fail to install for new users.

Producers also described receiving emails or other contacts from users with problems or feature requests, or saw questions come in specialist discussion forums.

> *I have created a Google group, a forum for people to ask questions about [package] [P-Gen-4]*

> *I would say that the bug reports are the most common type of contact. [S-Gov-1]*

Producers were motivated by personal contact with someone with a problem or idea, rather than on any systematic attempt to estimate the number of (potentially silent) users with different kinds of needs. One package author strongly emphasized this interest in particular issues rather than global information about the user base:

> *"Users needs" is a tricky thing in open source R packages...It is not something I will anticipate (I am not a commercial software vendor), nor do I care to. However, if a request is made which seems reasonable and not too time consuming, then I will usually oblige. [P-Cran-2]*

Most interviewees did not elaborate on which use cases they cared about in particular, although there were suggestions that they considered dealing with others' needs to be a cost, to be weighed against a benefit:

> *When you develop something like a package, it becomes almost a burden. Because suddenly, you have all these people who need your help with their specific little problem. And you wanna be able to help them because that means that your paper gets cited. [P-Gen-2]*

> *From time to time, I get requests of, "Can this be done? Can that be done?" And if I am not able to do the changes easily, then it goes pretty far down on the list of things. If it is something that actually could produce a paper or some collaboration, it would be more motivating to do this change [P-Cran-1]*

It could be that producers' apparent interest in addressing particular new use cases, rather than proactively counting kinds of users, is because a single, distinctive use case is more likely to be scientifically interesting, both to the producer as a collaborator, and to the rest of the community, resulting in a citation. In contrast, student educational use, as well as further examples of "typical" uses, may be less interesting and less likely to lead to new science and new citations.

But fixing other people's problems was also a moral obligation for some:

> *I do feel like, for me, it's a moral responsibility. I don't want there to be any mistakes in the code and if somebody's having trouble with a dataset, maybe there's a mistake and I need to find it. [P-Gen-2]*

**Producers' evaluation** Producers' reactions to the tool and their current practices suggest that software usage statistics would be most useful to them in understanding the needs of other users, if we provided them in a more granular form. When examining the tool as it exists, they imagined that they would use it to answer very specific questions about usage driven by tasks initiated for other reasons, for example to check how commonly people are using some method whose interface they wish to change.

One producer imagined how he could use USERPROJECTS to get more specific information about how his packages were being used:

*You could kind of get an idea of what they are using the package for. Many citations I get are more or less for a few functions, which seem to be comfortable to do in [my package] in comparison to another program. So, [you could] look up what they are using it for, maybe adjusting some model comparison or a very specific function. [P-Gen-3]*

Two producers mentioned wanting to be able to query USERPROJECTS for a particular abstract functionality, and see all the packages where that function was performed.

*I mean would it be absurd to actually work out which functions are used and which packages they belong to? [P-Gen-5]*

### Standards and duplication of effort

**Producers' practices** Several producers described doing careful research into the ecosystem's package offerings at the start of a new project, when deciding whether to add functionality to the ecosystem or relying on existing functionality. Interviewees searched, sometimes extensively, for related software to build on, rather than starting from the assumption that they would build everything from scratch.

Although they did search to see if functionality was already available before writing their own code, producers weren't averse to duplication of functionality if it made things more convenient. One author sought out implementations of two algorithms, Hierarchical F-statistic and AMOVA (Analysis of MOlecular VAriance), among phylogenetics packages. He found them in different packages, but realized it would be better for users if he re-implemented both:

*That's much easier for a user. ... They run one function, and they get one data frame result. If they have multiple packages then they have to combine them, then, say, the names of populations get changed, or the formatting is changed... It just becomes more work. [P-Gen-1]*

Coincidentally, a very similar situation came up for another producer involving the same statistic; in this case, the producer showed a preference for development effort over coordination effort:

*I was using Hierfstat to estimate FST but I had to rearrange my data in a way that Hierfstat would take it. And so, I ended up just writing my own code estimating FST so that I didn't have to rearrange my data and I didn't have to send [Hierfstat] this other patch. [P-Gen-2]*

**Stewards' practices** Keeping packages working together requires the work of software *producers*, but *stewards* set the ecosystem rules that determine what is expected of producers, how it will be enforced or encouraged, and how much work will be involved. In the case of CRAN, there is considerable cost to a producer of meeting the basic requirements of submitting a package. There is an R package called devtools [37] with the sole purpose of building and testing a submission of R code to evaluate it with respect to CRAN's requirements for compatibility with R itself and with neighboring packages; the instructions for using this package describe the process as "frustrating, but worthwhile", because "CRAN provides discoverability, ease of installation and a stamp of authenticity." [36]

This burden can be a disincentive to producers, as mentioned above.

*I should actually have some incentive of getting all my packages updated and better [...] but then it's a question of how to do it so that it don't break the code that other people wrote to use it. [P-Cran-1]*

All of the ecosystems inherited the basic technical coherence of the R platform, but stewards also echoed what producers told us about the importance of an ecosystem having shared standards, especially common classes:

*What we'd like to see is people reusing the fundamental objects and attitudes that we've sort of founded the project on. But there isn't a lot of enforcement there. There's a lot of recommendation. [S-Bio-2]*

Some stewards, like the core team of Bioconductor, actively nudge new contributors towards reuse of standard classes. rOpenGov and FLR both grew out of single packages, and so also have standard classes they can ask contributors to reuse. rOpenSci focuses on building community around common goals and themes, but S-Sci-1 also reported working towards cohesion:

*A number of our packages for interacting with Web services for scholarly articles: ... each of those has a slightly different programmatic interface. And so I've been working on a client that will integrate across all of those and so the user only has to learn one thing [S-Sci-1]*

Stewards thus face a tension between creating strict interoperation standards, thereby risking too much extra work for producers, on one hand, versus the risk of allowing incompatibilities to leak through to users of the ecosystem, on the other hand.

Producers addressing their own needs led to some duplication of functionality, and ecosystem stewards did not worry much about such duplication either:

*You can contribute a package; it goes through a certain amount of quality control. There isn't a lot of attention to semantic overlaps between functionalities. I'm sure there's some. [S-Bio-2]*

On the other hand there is a recognition of the value of reducing duplication by adopting standards: one steward described how the core group nudged their ecosystem to settle on one of two competing classes:

*It is possible that the group that has made the competing tool chain are still using that. You don't require them to use the [standard] one, but we have a lot more documentation and demonstration. And our fundamental location resources use the one that came out of the core, so that's sort of what drives the energy toward that particular solution. [S-Bio-2]*

**Producers' and stewards' evaluation** Producers saw some features of the Map as being useful for exploration of related packages when considering adding new functionality; this seems to be a key point where duplication can creep in if search is too difficult:

*There are packages there that I've never heard of. And it might show me that there's something else that I could use. [P-Gen-5]*

One producer imagined that he could look through USERPROJECTS to see how other projects were accomplishing a task, to see whether he should reuse some existing function, or implement something new:

*So should I have a wrapper in [my package] for something, say, that's in [a common package]? Or do I even need to have that function in [my package]? Can I just drop it because it's fully handled in [the common package]? [P-Gen-1]*

Stewards' ideas about how to use the Map suggested that it could help with their goal of nudging producers towards use of standard packages. In particular some stewards saw COUSAGE as useful for guiding documentation and user training development.

One steward told us he thought examples of how to use packages could be drawn from what COUSAGE reveals about the variety of contexts in which the package was used. In our interview, we showed him a package in his ecosystem, and he was surprised to see it often used with a set of packages he'd never heard of:

*If we knew people were using those, we could do tutorials with those, kind of demonstrate: this is how you do some sort of analysis to make a plot or whatever it is that people do with that [S-Sci-1]*

S-Gov-1 also saw COUSAGE as a way of gathering common use cases for training or improved documentation:

*We can also use that to think how to communicate the different projects. And maybe the documentation is more important for the more common projects. [S-Gov-1]*

**How data matched interviewees' expectations**
Interviewees questioned two aspects of the portrayal of their ecosystems: which "used-with" relationships were most important, and whether largely-duplicated projects were properly considered as "usage".

**COUSAGE** Generally, interviewees found the COUSAGE visualization informative once they understood it, but confusing at first. We struggled to find the right way to prune this visualization to a reasonably "important" neighborhood of a package, and in some cases the results were not intuitive. In particular, we used a pointwise mutual information measure to prioritize relationships, but among Bioconductor packages, this resulted in some visualizations showing very low-frequency neighbors:

*It's pretty clear that you're working with people that have very specific focuses... I don't know that much about it, but this package is definitely not going to be one of the dominant users of BiocGenerics. [S-Bio-2]*

S-Bio-1 suggested that the graph should be viewer-dependent, emphasizing connections to very domain-specific packages useful for scripting when displaying the graph to an end user, and favoring internal utility packages when displaying to producers of packages.

Some were surprised not to see things they expected:

*Yeah, it makes sense. I'm somewhat surprised we don't see the other rOpenGov packages. Because I think they are probably used with Sorvi also. At least I have been using them with Sorvi. [S-Gov-1]*

A general limitation of the data set we used which was most apparent in this visualization was that for rarely used packages (or, rarely used in Github scripts), the relationships were drawn from a very small sample of uses. A small ecosystem steward phrased this positively:

*The more commonly the package is used, the more useful this is. And I think it's also good to start to collect this early on. [S-Gov-1]*

**Inclusion of usage data** Interviewees were sometimes concerned about automatically duplicated projects distorting the dataset.

USERPROJECTS included "forks" (copies of projects, possibly modified); this caused some interviewees to ask if the counts in PACKAGELIST also counted forks (they did); they felt these should not be included. We had expected that a highly-forked project would be welcome evidence of more usage, but interviewees seemed to consider forks irrelevant noise.

Another interviewee drew our attention to a distinction between scientific use and educational use. In our examination of the data, school assignments in Github were often quite repetitive, with many students' work showing up as having similar filenames and imports. One steward worried that school assignments were being counted. These rarely showed up in our walkthroughs with interviewees, however, since the most highly forked Github R school assignments we examined happened to almost never use any R packages. However this could be a concern in the future if other sources of usage data tap into data from course assignments that do encourage students to use packages.

One producer raised the question of the time span over which usage data was aggregated: P-Cran-1 at first doubted the accuracy of PACKAGELIST because a new, highly efficient package appeared so far down the list. This suggests that an option to show a more recent "window" of usage could better depict the ecosystem's current status.

Some stewards also suggested that different data sources are appropriate for different communities. S-Flr-1 thought Github was an appropriate source because they encourage their users to use Github; but S-Sci-1 thought Github users might be atypical of rOpenSci users.

## DISCUSSION

### Tradeoff of development effort and coordination costs

Producers repeatedly described adding whole packages, and incremental functionalities, simply because it was inconvenient to reuse existing functionalities. Rather than trying to build consensus with authors of existing packages, they preferred to duplicate effort.

The tolerance for duplication is less surprising when considered in terms of the way scientists in this ecosystem think of costs and benefits. The benefit that they are trying to maximize is scientific impact, and it is often *distinctive use*, not repetitive use, that is more likely to contribute to novel, innovative results. So perhaps it is not surprising that scientists lean towards reimplementation of a function to get exactly the functionality they want, as opposed to making do with existing offerings.

As for the costs, our evidence indicates that the costs of coordinating with others often outweigh costs of developing lines of code. Prior research suggests that two factors in R ecosystems make coordination particularly expensive: First, many R producers are scientists, and Ribes et al. [30] point out that coordination among scientists can be difficult because of misalignment of their time schedules for research. Secondly, CRAN asks authors to test all dependencies with the latest versions of other packages when submitting updates to the repository [26] with every update. A failed test risks having their package dropped from the repository. This makes unresponsive collaborators a risk, compared with an ecosystem like Node.js in which packages can safely refer to previous, known versions of other packages, and update on their own schedule [4].

## IMPLICATIONS FOR DESIGN

Overviews of relationships between packages are useful for stewards, and for deliberate efforts like hackathons; but for common development tasks, producers need (1) concrete, granular information to reduce coordination cost: for example exactly which functions are being used by which other packages and how, and (2) motivating information about distinctive use cases and patterns, rather than only raw usage: for example perhaps a map whose elements are distinct **configurations** (collections of packages used together, like the *persona* glyphs described by Terry et al [32], clustered to show fields of similar use cases. (3) Ways of sorting and filtering by *functionality*, for example searching for methods and classes by name or words in the documentation. This would help with the duplication of effort problem, by making good information available at the key moment in time when a would-be package author decides whether to start a new package or extend an old one.

For stewards, the breadth of sources they drew on in characterizing ecosystem usage suggests many possible new data sources for extending the Map: counting and summarizing references to packages in social media such as email lists and twitter; ways of extracting textual references in blog posts, academic papers, grey literature, etc; statistics from visits, searches, and downloads from websites controlled by the ecosystem; and ways for core ecosystem people to mine their own email and contact software to characterize the number and volume of people they interact with about packages.

## CONCLUSION

In this paper we have introduced the Scientific Software Network Map, a tool designed to provide price-like signals of usage to participants scientific software ecosystems, to help them justify, motivate, and direct software work for the benefit of users of the software. We evaluated how the Map might help ecosystem participants, and gathered information about their current practices:

- Providing summaries of usage as a proxy for "price" in a software ecosystem has potential to help with a variety of tasks, but the signals must be chosen carefully. Producers' practices appear to work to maximize diversity of use cases and minimize coordination costs, so usage statistics that count duplicate uses may be less motivating than ones that count only distinct uses.
- Ecosystem *stewards* already present a variety of statistics about usage to justify funding for ecosystems; the Map could provide welcome support for collecting, aggregating, and presenting these price-like signals.
- Although the R ecosystem decomposes naturally at the level of "packages", for whom individual maintainers are held responsible, for the purpose of understanding user needs, usage statistics at a finer level of detail: particular functionalities, data structures and interfaces, could help producers decide which parts of their package it is safe to change.

These insights should lead to better information tools for scientific software communities, and we hope that these better tools in turn continue to reinforce the powerful magnifying effect that software has on science.

## ACKNOWLEDGMENTS

This material is based in part upon work supported by (redacted for review)

## REFERENCES

1. Anon. RStudio. Retrieved May 22, 2015 from http://www.rstudio.com/
2. D.E. Atkins et al. 2003. Revolutionizing Science and Engineering Through Cyberinfrastructure: Report of the National Science Foundation Blue-Ribbon Advisory Panel on Cyberinfrastructure.
3. Matthew J. Bietz and Charlotte P. Lee. 2012. Adapting cyberinfrastructure to new science. *iConference*, 183–190. http://dx.doi.org/10.1145/2132176.2132200
4. Christopher Bogart, Christian Kästner, and James Herbsleb. 2015 (to appear). When it breaks, it breaks: How ecosystem developers reason about the stability of dependencies. *Workshop on Softw. Support for Collab. Global Software Eng.*
5. Ronald F. Boisvert and Ping Tak Peter Tang. 2001. *The Architecture of Scientific Software* R. F. Boisvert and P. T. P. Tang, eds. Kluwer Academic Press, Boston.
6. Jeffrey C. Carver, Richard P. Kendall, Susan E. Squires, and Douglass E. Post. 2007. Software Development Environments for Scientific and Engineering Software: A Series of Case Studies. *Int. Conf. Softw. Eng.*, 550–559. http://dx.doi.org/10.1109/ICSE.2007.77
7. Laura Dabbish, Colleen Stuart, Jason Tsay, and Jim Herbsleb. 2012. Social Coding in GitHub: Transparency and Collaboration in an Open Software Repository. In *Proc. Conf. Computer Supported Cooperative Work (CSCW)*, 1277–1286.
8. Andrea Devenow and Ivo Welch. 1996. Rational herding in financial economics. *Eur. Econ. Rev.* 40, 603–615.
9. Yvonne Dittrich. 2014. Software engineering beyond the project - Sustaining software ecosystems. *Inf. Softw. Technol.* 56, 11, 1436–1456. http://dx.doi.org/10.1016/j.infsof.2014.02.012
10. Sebastian Draxler and Gunnar Stevens. 2011. Supporting the collaborative appropriation of an open software ecosystem. *Comput. Support. Coop. Work (CSCW)* 20, 4-5, 403–448. http://dx.doi.org/10.1007/s10606-011-9148-9
11. Georgios Gousios. 2013. The GHTorent dataset and tool suite. *IEEE Int. Work. Conf. Min. Softw. Repos*, 233–236. http://dx.doi.org/10.1109/MSR.2013.6624034
12. Nicole Haenni, Mircea Lungu, Niko Schwarz, and Oscar Nierstrasz. 2014. A Quantitative Analysis of Developer Information Needs in Software Ecosystems. *European Conference on Software Architecture Workshops (ECSAW)*.
13. James Howison and Julia Bullard. In Press: Software in the Scientific Literature: Problems with Seeing, Finding, and Using Software Mentioned in the Biology Literature. *J. Assoc. Informait. Sci. Technol.* http://dx.doi.org/10.1002/asi.23538
14. James Howison, Ewa Deelman, Michael J. Mclennan, Rafael Ferreira, and James D. Herbsleb. In press. Understanding the scientific software ecosystem and its impact: current and future measures. *Research Evaluation.* http://rev.oxfordjournals.org/cgi/reprint/rvv014?ijkey=TfzJc5bI7X5Xk0v&keytype=ref
15. James Howison and James D. Herbsleb. 2011. Scientific software production: incentives and collaboration. *Proc. Conf. Comput. Support. Collab. Work*, 513–522.
16. James Howison and James D. Herbsleb. 2013. Incentives and integration in scientific software production. *Proc. Conf. Comput. Support. Coop. Work (CSCW),* 459–470. http://dx.doi.org/10.1145/2441776.2441828
17. Steven J. Jackson, David Ribes, Ayse G. Buyuktur, and Geoffrey C. Bowker. 2011. Collaborative Rhythm: Temporal Dissonance and Alignment in Collaborative Scientific Work. *Proc. Conf. Comput. Support. Collab. Work (CSCW)*, 245–254.
18. Slinger Jansen, Anthony Finkelstein, and Sjaak Brinkkemper. 2009. A Sense of Community: A Research Agenda for Software Ecosystems. In *Int. Conf. Software Engineering (ICSE) -- Companion Volume*, 187–190.
19. Daniel S. Katz et al. 2014. Summary of the First Workshop on Sustainable Software for Science: Practice and Experiences (WSSSPE1). *J. Open Res. Softw.* 2, 1, e6: 1–21.
20. Robert E. Kraut and Paul Resnick. 2012. *Building Successful Online Communities: Evidence-Based Social Design*. MIT Press, Cambridge, MA.
21. Karim R. Lakhani and Robert G. Wolf. 2003. *Why Hackers Do What They Do: Understanding Motivation Effort in Free/Open Source Software Projects*.
22. Hilmar Lapp. Population Genetics in R Hackathon. Retrieved May 22, 2015 from https://github.com/NESCent/r-popgen-hackathon

23. C. Lee, M. Bietz, and D. Ribes. 2008. Designing cyberinfrastructure to support science. *Comput. Support. Coop. Work Conf. Workshops (CSCW)*.
24. Charlotte P. Lee, Paul Dourish, and Gloria Mark. 2006. The human infrastructure of cyberinfrastructure. *Comput. Support. Coop. Work (CSCW)*, 483–492. http://dx.doi.org/10.1145/1180875.1180950
25. Mircea Lungu, Michele Lanza, Tudor Girba, and Romain Robbes. 2010. The Small Project Observatory. *Sci. Comput. Program.* 75, 264–275. http://dx.doi.org/10.1016/j.scico.2009.09.004
26. CRAN Repository Maintainers. CRAN Repository Policy. Retrieved September 25, 2015 from https://cran.r-project.org/web/packages/policies.html
27. Konstantinos Manikas and Klaus Marius Hansen. 2013. Software ecosystems-A systematic literature review. *J. Syst. Softw.* 86, 5, 1294–1306. http://dx.doi.org/10.1016/j.jss.2012.12.026
28. Audris Mockus, Roy T. Fielding, and James D. Herbsleb. 2002. Two case studies of open source software development: Apache and Mozilla. *ACM Trans. Softw. Eng. Methodol.* 11, 3, 309–346. http://dx.doi.org/10.1145/567793.567795
29. W. Powell. 1990. Neither Market nor Hierarchy: Network Forms of Organization. *Res. Organ. Behav.* 12, 295 – 336. http://dx.doi.org/10.1590/S1415-65552003000200016
30. David Ribes and Thomas A. Finholt. 2009. The Long Now of Technology Infrastructure: Articulating Tensions in Development. *J. Assoc. Inf. Syst.* 10, 5, Article 2, 375–398.
31. Scopus. Elsevier. Retrieved May 22, 2015 from http://www.scopus.com
32. Michael Terry, Matthew Kay, Brad Van Vugt, Brandon Slack, and Terry Park. 2008. ingimp: Introducing Instrumentation to an End-User Open Source Application. *Conf. Hum. Factors Comput. Syst.*, 607–616.
33. Erik Trainer, Chalalai Chaihirunkam, Arun Kalyanasundaram, and James Herbsleb. 2015. From personal tool to community resource: what's the extra work and who will do it? *Proc. Conf. Comput. Support. Collab. Work*, 417–430.
34. Jason Tsay, Laura Dabbish, and James Herbsleb. 2014. Influence in github. *Int. Conf. Softw. Eng.*, 356–366. http://dx.doi.org/10.1145/2568225.2568315
35. Patrick A. Wagstrom. 2009. *Vertical Interaction in Open Software Engineering Communities. DTIC Report.*
36. Hadley Wickham. 2015. *R packages*. O'Reilly Media.
37. Hadley Wickham and Winston Chang. 2015. devtools: Tools to Make Developing R Packages Easier.
38. Yu Wu, Jessica Kropczynski, Patrick C. Shih, and John M. Carroll. 2014. Exploring the ecosystem of software developers on GitHub and other platforms. *Proc. companion Publ. Conf. Comput. Support. Coop. Work Soc. Comput. (CSCW Companion)*, 265–268. http://dx.doi.org/10.1145/2556420.2556483